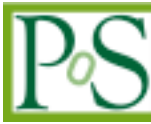

# LOFAR, E -LOFAR and Low-Frequency VLBI

**M.A. Garrett[1][a][b][c], H. Rampadarath [b][d], E. Lenc [e], O. Wucknitz [f]**

[a]*ASTRON, Netherlands Institute for Radio Astronomy PO Box 2, 7990AA Dwingeloo, The Netherlands*
[b]*Sterrewacht Leiden, Postbus 9513, 2300 RA Leiden, The Netherlands*
[c]*Centre for Astrophysics and Supercomputing, Swinburne University of Technology, Australia*
[d]*Joint Institute for VLBI in Europe, PO Box 2, 7990AA Dwingeloo, The Netherlands*
[e]*Australia Telescope National Facility, CSIRO, Epping NSW 1710, Australia*
[f]*Argelander-Institut für Astronomie, Universität Bonn, Auf dem Hügel 71, 53121, Bonn, Germany.*

*E-mail:* `garrett@astron.nl`

The Low Frequency Array (LOFAR) is a new generation of electronic radio telescope based on aperture array technology. The telescope is being developed by ASTRON, and currently being rolled out across the Netherlands and other countries in Europe. We present the current status of the project, and its relation to high resolution instruments such as the European VLBI Network (EVN) and the Very Long Baseline Array (VLBA). In particular, recent VLBI results at 327 MHz associated with: (i) a shallow survey based on VLBA archive data and (ii) a deep, wide-field Global VLBI survey centred on two in-beam calibrators, B0218+357 and J0226+3421 are discussed. The results suggest that there will be no shortage of relatively bright primary calibrators that remain unresolved by LOFAR even on the longest European baselines. The sky density of fainter "in-beam" calibrators should also be more than adequate to permit the generation of high fidelity images over a large fraction of the sky, especially in the high-band observing band (120-240 MHz). Extending LOFAR via international stations to baseline lengths of several thousand kilometres is certainly practical and should significantly enhance the scientific output and capabilities of the array.

**1. Introduction**

ASTRON, the Netherlands Institute for Radio Astronomy, is developing and building LOFAR (the Low Frequency Array) – an ambitious, next generation radio telescope that fully exploits recent advances in digital signal processing, fibre-based distributed communication networks and high performance super-computing. Operating in a largely unexplored region of the electro-magnetic spectrum (15-240 MHz), LOFAR will consist of a distributed interferometric array of dipole antenna stations that have no moving parts, and that permit large areas of the sky to be imaged simultaneously – up to 8 independent fields of view can be instantaneously steered to any position on the sky using electronic beam-forming techniques. Over the next 2 years, 40 stations are expected to be built in various locations across the Netherlands. The densely populated core comprising 13 antennas is located in Exloo, not far from ASTRON's headquarters in Dwingeloo. It is expected that up to 40 LOFAR stations will be located in the Netherlands, and at least an additional 10 antennas will be located in other European countries (Germany, UK, Sweden, France & Italy). Two types of antenna are used: the low-band antenna (LBA, 15-80 MHz) and the high-band antenna (HBA 110-240 MHz).

---
[1] Speaker

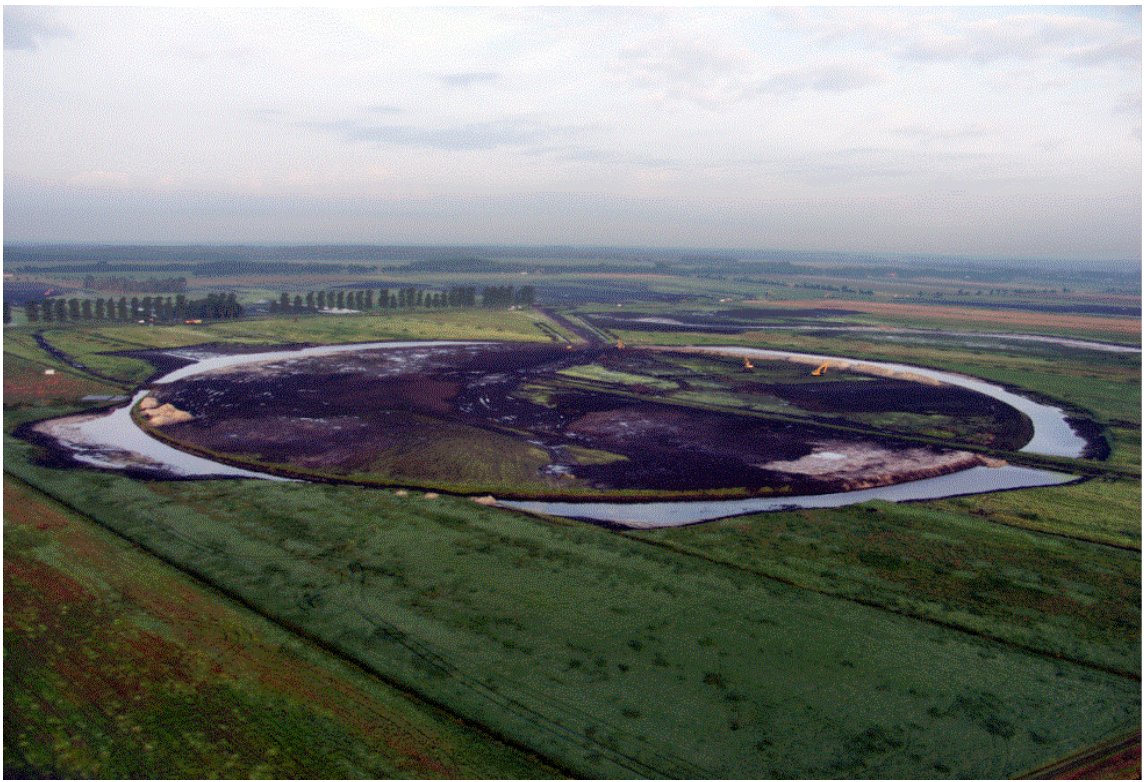

*Figure 1: the LOFAR " super-terp"under construction and as it appeared around the time of the EVN IX Symposium (September 2008). The "super-terp" is raised ~ 1 metre above the local ground level and measures 360 metres across. Six of the thirteen LOFAR core stations will be located here. The other 7 core stations are located within a few km of the super-terp.*

LOFAR represents a step-change in the evolution of radio astronomy technology. It offers many new modes of observation, including an all-sky transient detection capability. By buffering large amounts of data at the individual dipole level, retrospective imaging of the entire sky is also possible. LOFAR's ability to look back in time, together with the telescopes huge field of view, ushers in a new era for radio astronomy. In addition to the study of transient phenomena, the LOFAR Key Science Programmes (KSP) include: (i) deep, wide-field extra-galactic surveys; (ii) exploring the epoch of reionisation, and (iii) cosmic ray detection (see www.lofar.org for more information). When completed in 2010, the full LOFAR array will have a sensitivity and resolution that is ~ 2 orders of magnitude better than anything that has proceeded it, and will open up one of the last unexplored frontiers of observation astronomy.

**2. LOFAR system**

The Dutch LOFAR stations are actively populated with ~ 48 dual polarisation LBA elements and 48 HBA tiles, (each HBA tile is composed of 16 dual-polarisation antenna elements). In fact at each station, 96 LBA elements are physically present, but only 48 are active at any one time – the exact configuration employed depends on the observing frequency. A typical station spans about 180 x 100 metres across. To ensure good sensitivity on the longer baselines, the international stations have twice as many antenna elements as the Dutch stations (see de Vos et al. 2009 for a complete system description).

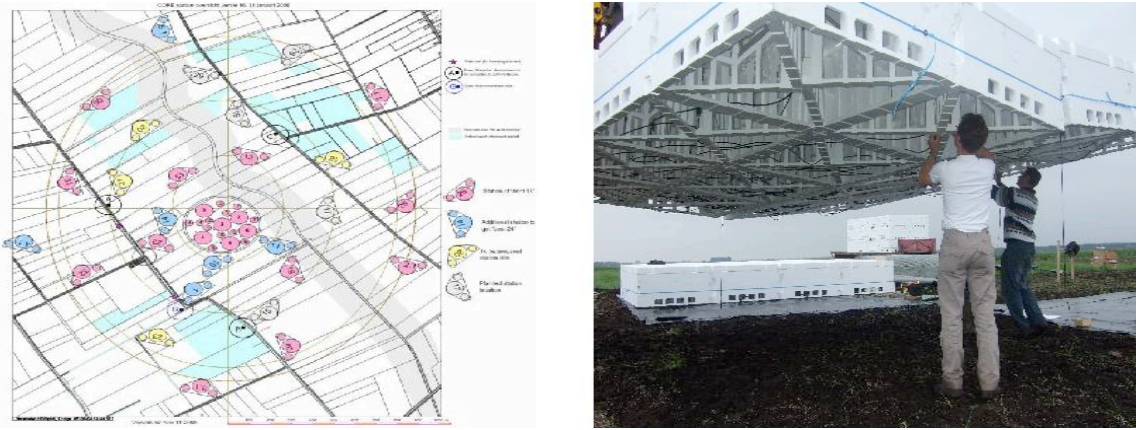

*Figure 2: The location of the LOFAR core stations at Exloo (left). The layout of each individual core stations resembles "Micky Mouse"a little, with the 96 low-band antennas being flancked by 2 x 24 high band antenna tiles. An image of the prototype HBA tiles being placed in Exloo is shown (right).*

LOFAR includes a core area about 2 km across that includes 13 stations (see Figure 1 and 2). A raised area located in Exloo, and known as a "superterp" (in Dutch), is 360-metre across, and accomodates 6 of the core stations closely packed together. This ensures good (short-spacing) uv-coverage that is important to several of the KSPs. Stations in the core have a somewhat different configuration to those located outside of the core. In particular, the core stations have the HBA elements split into 2 lots of 24 tiles. The entire telescope (including international stations) is remotely controlled from ASTRON's Radio Observatory Control Centre, located in Dwingeloo, the Netherlands.

The data associated with each dipole in a station are coherently combined together, sampled and digitised using up to 12-bit data representation. Eight independent, dual-polarisation beams are formed, each with a bandwidth of ~ 4 MHz and the associated data (~ 2.1 Gbps per station) are transferred via optical fibres to the correlator in Groningen (currently an IBM BlueGene/P super computer). Due to the huge data flows involved, the data are correlated and calibrated in real-time. Most astronomers are expected to interact with the data via calibrated images, accessed via a large online database. Access to (averaged) visibility data will also be possible.

In July 2007, the LOFAR design underwent a significant de-scope. An updated summary of the main characteristics of the telescope are provided in Table 1 (for more details see [1]). Note that it is possible to trade beams for bandwidth: if only 1 beam is formed per station, a total bandwidth of 32 MHz can be employed (improving the quoted noise figure in table 1 by ~ 2-3).

| Freq (MHz) | Field of View (Sq. degrees) | Resolution LOFAR-NL (arcsecs) | Resolution International LOFAR (arcsecs) | LOFAR remote station SEFD (kJy) | LOFAR-NL 1-$\sigma$ image noise level (mJy) |
|---|---|---|---|---|---|
| 30 | 65.8 | 16 | 1 | 89 | 20 |
| 60 | 105 | 8 | 0.5 | 32 | 7 |
| 120 | 16 | 4 | 0.25 | 1.8 | 0.4 |
| 180 | 7 | 3 | 0.20 | 1.6 | 0.4 |
| 240 | 4 | 2 | 0.15 | 2.0 | 0.5 |

*Table 1: The main characteristics of LOFAR at a few representative frequencies. Note that the field-of-view of the LBA stations is dependent on details of the dipole configuration which changes at 40 MHz. The 1-sigma r.m.s. noise level assumes an integration time of 1 hour, and 1 beam with 2 polarisations, 4 MHz bandwidth (see [1] for a full description).*

### 3.0 International LOFAR: challenges ahead

LOFAR stations will be located in various countries in Europe. Currently at least 5 stations are planned in Germany, 4 in the UK, 1 in Sweden and 1 in Italy. Several other countries are currently seeking funding for additional European stations (e.g. Poland, Ukraine/Austria and Ireland). Figure 3 shows the overall configuration of the LOFAR telescope. The first international (LBA) LOFAR station (see Figure 4) is already complete, and is located at the MPIfR, Effelsberg, alongside the 100-metre telescope. The international stations will also be connected to the LOFAR correlator in Groningen, building on the techniques pioneered by e-VLBI (see Szomoru et al. these proceedings). With baseline lengths extending beyond 1500 km, these European stations will improve the resolution of LOFAR by more than an order of magnitude. The extra collecting area provided by the international stations will also enhance the point-source sensitivity of the array. For example, the numbers quoted in table 1 can be improved (i.e. reduced) by a factor of ~ 0.7. The Survey KSP in particular, will greatly benefit from the enhanced capabilities of the telescope.

The extension of LOFAR to baseline lengths of a few thousand kilometres presents several new challenges. With a resolution of up to 0.1 arcseconds, the sky density of suitable calibrator sources may be a significant issue, especially at the lowest frequencies. In addition, very little is known about the morphology of radio sources at these low frequencies and high resolutions. To address some of these issues, we have used the VLBA and Global VLBI to study the low-frequency radio sky at the highest angular resolution.

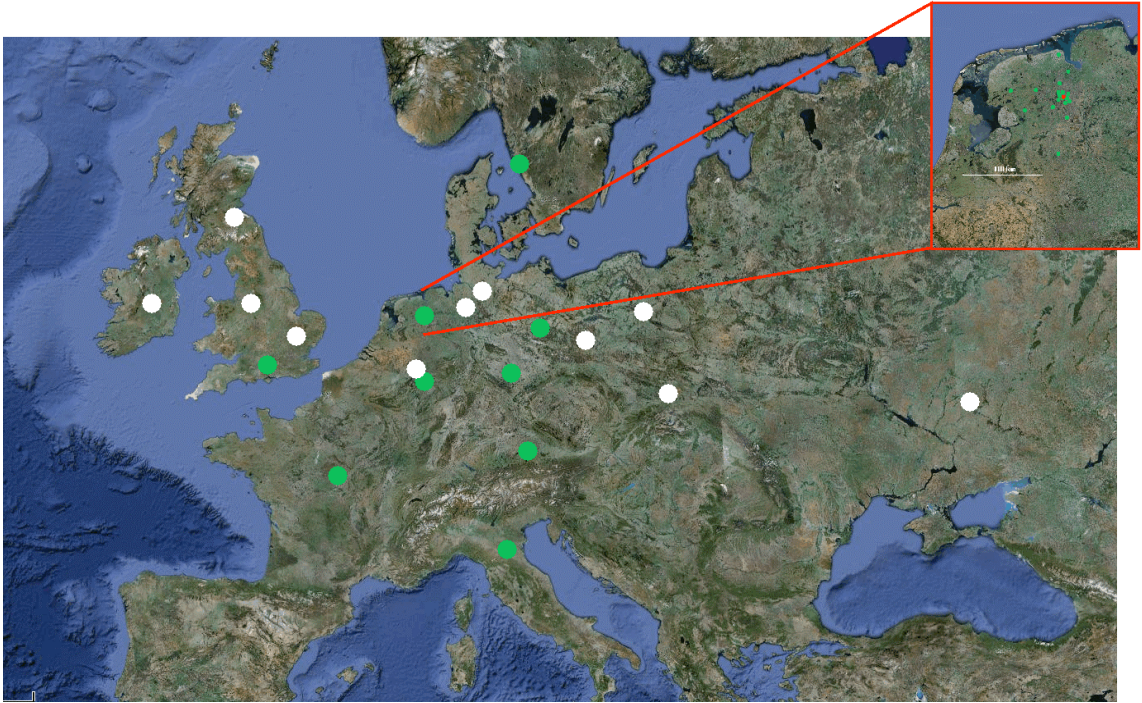

*Figure 3: The location of LOFAR stations in Europe and in the Netherlands. Green dots represent stations where funding is secured or for which letters of intent have been received. White dots represent stations that are planned by the various international consortia and for which funding is being sought. The location of remote antennas in the Netherlands is shown top right – the red dot shows the location of the central core stations in Exloo*

## 3. Low-frequency VLBI

The VLBA and a subset of the EVN antennas regularly conduct VLBI observations at frequencies that are not so far removed from the top-end of the LOFAR HBA range (240 MHz). While most VLBI observations take place at centimetre wavelengths (where the sky noise is low and the coherence times are relatively long), VLBI observations at 327 MHz are common and essentially routine. A trawl through the literature however, reveals a disappointing number of refereed or even conference publications. We have embarked on two programmes: (i) the automatic analysis of 43 sources drawn from the VLBA archive observed at 327 MHz over the period 2003-2006 (Rampadarath, Garrett & Polatidis 2009 submitted), and (ii) a deep, wide-field 327 MHz VLBI survey of 272 sources located within the field of B0218+357 (Lenc et al. 2008).

**3.1 VLBA 327MHz archive survey**

We have analysed VLBA 327 MHz archive data of 43 extragalactic sources in order to identify early targets and primary calibrator sources for LOFAR[2]. Some of these sources will also be suitable as ``in-beam'' calibrators, permitting deep, VLBI wide-field studies of other faint sources in the same field of view (see section 2.2). The data were observed between 1 January 2003 to December 31 2006, and have been analysed via an automatic pipeline, implemented within AIPS.

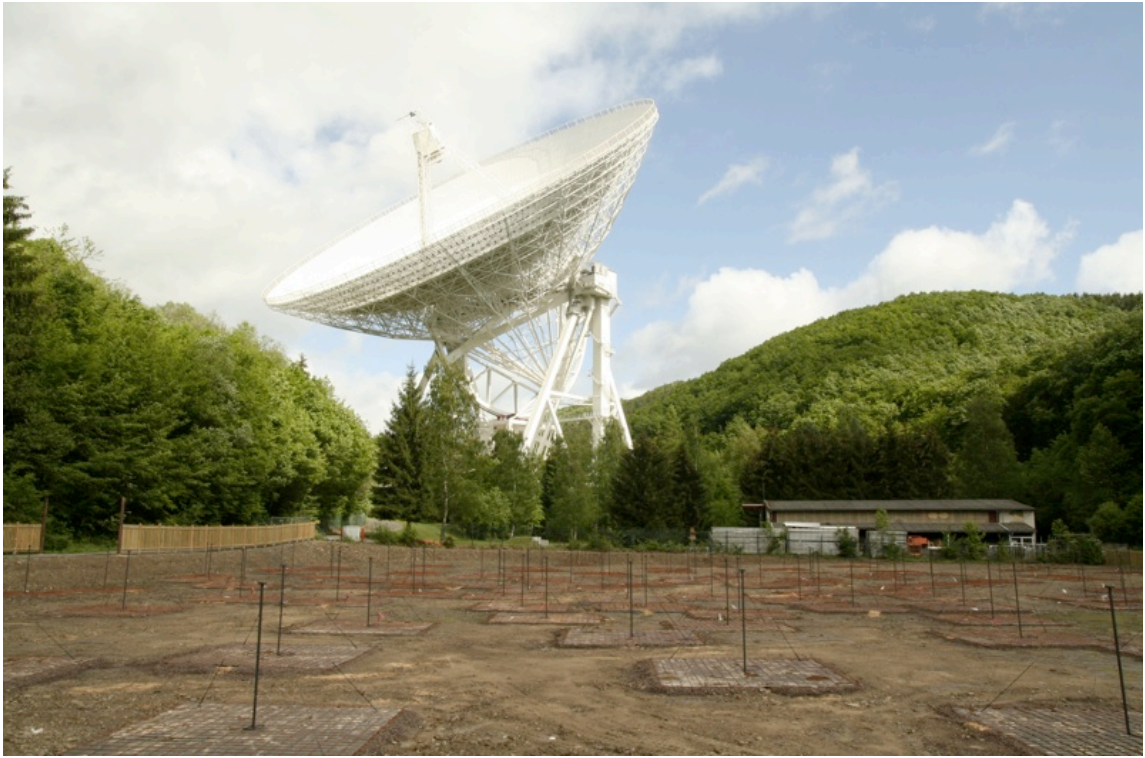

*Figure 4: the first international LOFAR LBA station is located at Effelsberg, MPIfR.*

The vast majority of the data are unpublished. The sample consists of 43 sources, of which 40 have been detected by us on at least some limited sub-set of VLBA baselines. Twenty nine are bright enough and have sufficient data to be successfully imaged. Most of the sources are compact, but a significant fraction also show resolved, extended structures (see Figure 5). Of the 29 sources that were imaged, 13 were detected on the longest VLBA baselines (9 Mega-wavelengths), the remainder were detected on baselines greater than 2 Mega-wavelengths (corresponding to the maximum baseline of LOFAR including the international baselines). Potential primary amplitude calibrators (i.e. sources likely to be unresolved by E-LOFAR) include: 3C84, 0402+379, 3C111, J0423-0120, 0537-286, 1148-001, 3C345, J1816+3457, 2007+777, J2022+6136, BL Lac, 3C454.3, J2327+0940, J2330+1100. Many of these sources might also be sensibile targets for the RadioAstron space VLBI mission.

Figure 5 shows that radio sources that dominate continuum images produced by the low-resolution LOFAR test station in Exloo, can appear very differently at (roughly) the same radio frequency but when observed at much higher angular resolution. While Virgo-A (3C274) shows evidence for collimated AGN jet emission on milliarcsecond scales (also Figure 5), Cyg-A goes entirely undetected. Sources that are suitable as calibrators for the short-baseline Netherlands array may not be suitable for international baselines.

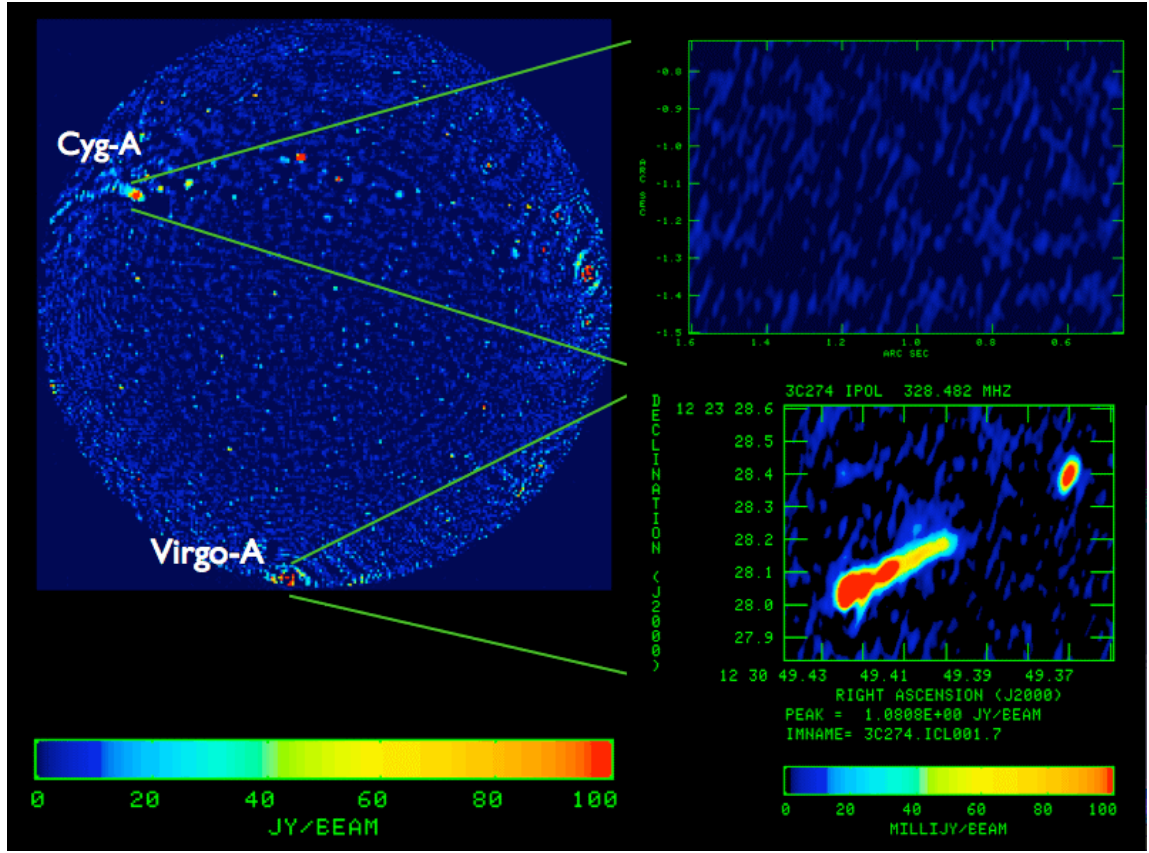

*Figure 5: Top left – an all-sky LOFAR image (courtesy Sarod Yatawatta) shows two of the brightest sources in the low-frequency radio sky – Cyg-A and Virgo-A. While many of these sources will be heavily resolved by the international LOFAR stations (e.g. Cyg-A – top right), others will remain compact (e.g. Virgo-A – bottom right). In particular, the presence of many fainter but compact sources within each LOFAR beam, should permit a large fraction of the sky to be mapped out with high angular resolution and good mage fidelity [2,3].*

**3.2 A Deep, high-Resolution, wide-field 327 MHz VLBI Survey**

We have conducted the first wide-field, very long baseline interferometry (VLBI) survey at 327 MHz. The survey area consists of two over-lapping 28 square degree fields centered on the quasar J0226+3421 and the gravitational lens B0218+357. A total of 272 sources were targeted in these fields, based on the positon of known sources drawn from the Westerbork Northern Sky Survey (WENSS) and the sensitivity limit of the VLBI observations. A total of 27 sources were detected as far as 2° from the phase centre (a subset of the detections are shown in Figure 6).

The results of the survey suggest that at least 10% of moderately faint (S~100 mJy) sources found at 90 cm contain compact components smaller than ~0.1"-0.3" and stronger than 10% of their total flux densities. This survey is the first systematic (and non-biased), deep, high-resolution survey of the low-frequency radio sky. It is also the widest field of view VLBI survey with a single pointing to date, exceeding the total survey area of previous higher frequency surveys by 2 orders of magnitude. These initial results suggest that new low-frequency telescopes, such as LOFAR, should detect many compact radio sources and that plans to extend these arrays to baselines of several thousand kilometers are warranted. We also note that efficient algoithms are being developed (Wucknitz in prep) that are able to image the full primary with VLBI resolution. Our wide-field 327 MHz VLBI data have been used as a test case, to produce an image that covers an area of 4 square degrees on the sky and occupies $256000^2$ pixels.

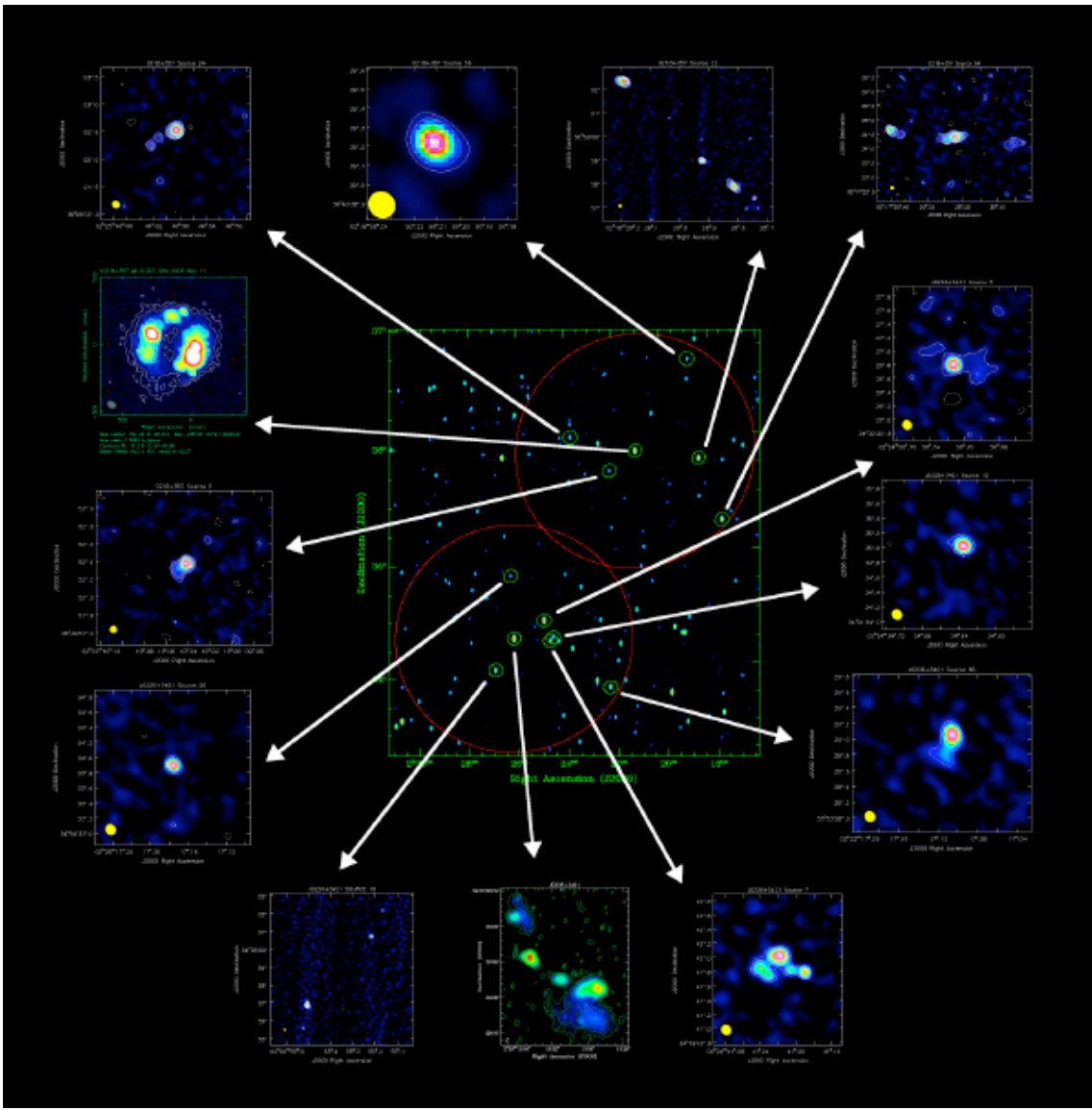

*Figure 6: Global VLBI images of about half of the sources detected in the deep, wide-field 327MHz survey[3].*

**Acknowledgements**

LOFAR is funded by the Dutch government through the BSIK programme for interdisciplinary research and improvement of the knowledge infrastructure. Additional funding is provided through the European Regional Development Fund (EFRO) and the innovation programme EZ/KOMPAS of the Collaboration of the Northern Provinces (SNN). ASTRON is an institute of the Netherlands Organisation for Scientific Research (NWO). O.W. is funded by the Emmy-Noether-Programme of the Deutsche Forschungsgemeinschaft, reference WU588/1-1.